\documentclass[12pt, a4paper]{article}
\pdfoutput=1

\usepackage{amsfonts}
\usepackage{amssymb, amsmath, bm}
\usepackage{graphicx, rotating}
\usepackage{epsfig, epstopdf}
\usepackage{array}
\usepackage{latexsym}
\usepackage{xcolor}
\usepackage{cite}
\usepackage{slashed, cancel}
\usepackage{hyperref}
\usepackage[normalem]{ulem}
\usepackage[utf8]{inputenc}
\usepackage[export]{adjustbox}
\usepackage{multirow, verbatim}

\newcommand{\be}{\begin{equation}}
\newcommand{\ee}{\end{equation}}
\newcommand{\bea}{\begin{eqnarray}}
\newcommand{\eea}{\end{eqnarray}}
\newcommand{\nn}{\nonumber}

\begin{document}

\begin{titlepage}

\begin{flushright}
\small
DESY 21-089
\end{flushright}
\vspace{.3in}

\begin{center}
{\Large\bf
On the wall velocity dependence \\ of electroweak baryogenesis \vskip 0.2 cm}%
\bigskip\color{black}
\vspace{1cm}{
  {\large
Glauber C. Dorsch$^a$,
Stephan~J.~Huber$^b$,
Thomas Konstandin$^c$
}}

{\small\vskip5mm
$^a$ Departamento de F\'isica, UFMG,  31270-901, Belo Horizonte, MG, Brazil \\
$^b$ University of Sussex, Brighton, BN1 9QH, UK \\
$^c$ DESY, Notkestra{\ss}e 85, 22607 Hamburg, Germany\\
}

\bigskip

\begin{abstract}
We re-evaluate the status of supersonic electroweak baryogenesis using a generalized fluid Ansatz for the non-equilibrium distribution functions. Instead of truncating the expansion to first order in momentum, we allow for higher order terms as well, including up to 21 fluctuations. The collision terms are computed analytically at leading-log accuracy. We also point out inconsistencies in the standard treatments of transport
in electroweak baryogenesis, arguing that one cannot do without specifying an Ansatz for the distribution function. We present the first analysis of baryogenesis using the fluid approximation to higher orders. Our results support the recent findings that baryogenesis may indeed be possible even in the presence of supersonic wall velocities.
\end{abstract}

\end{center}

\end{titlepage}

\newpage
\section{Introduction}
\label{sec:Introduction}

Despite the immense empirical successes of the Standard Model, the origin of the matter-antimatter asymmetry in the Universe remains one of the outstanding open problems in particle physics. The issue is particularly tantalizing because our current theories do seem to contain all the necessary ingredients for solving the problem, namely baryon number non-conservation, charge and charge-parity violation, and non-equilibrium dynamics. Indeed, in the Standard Model baryon number is violated by processes which become very efficient at high temperatures, such as those found in the early Universe, where one also has non-equilibrium dynamics such as the Hubble expansion and possibly phase transitions, while C and CP violation are present in the weak and fermionic sectors. However, the requirement that these ingredients appear simultaneously and in a sufficient amount turns out to be a severe constraint, making it difficult for a concrete mechanism to be devised that correctly predicts the observed asymmetry.

A very appealing approach is \emph{non-local electroweak baryogenesis}~\cite{Nelson:1991ab, Morrissey:2012db, Konstandin:2013caa}, which has attracted much attention from the community in the past decades for its phenomenological and cosmological implication at current and near-future experiments. The mechanism relies on charge transport along bubbles of the Higgs field in a first-order electroweak phase transition in the early Universe. The basic idea is that particle collisions in front of the expanding bubble may lead to a net CP asymmetry, which can be converted to a baryonic asymmetry by baryon-number-violating sphaleron processes in front of the wall. When the asymmetry is diffused into the bubble, where the sphalerons are inefficient, the reverse baryon-wash-out reactions are suppressed, ensuring that the Universe remains asymmetric to the present day.

Because this mechanism relies on diffusion of particles in the primordial plasma, it has been common lore that it could not work for supersonic wall speeds, since in this case the time for non-equilibrium processes to take place in the plasma before the bubble sweeps past would supposedly be insufficient. This intuition was confirmed in the first works that 
studied the velocity dependence of electroweak baryogenesis in detail~\cite{Fromme:2006wx}.

However, this claim has been disputed in a recent work by Cline and Kainulainen~\cite{Cline:2020jre}. The main argument is that diffusion, being an essentially microscopic non-equilibrium process, heavily dependent on particle interactions, is not necessarily related to a collective phenomenon such as the propagation of sound in the plasma. Indeed, the Bose-Einstein and Fermi-Dirac equilibrium distribution functions show no regard for the sound speed $c_s\simeq 1/\sqrt{3}$, and some of the particles do have larger velocities and could diffuse into a supersonic bubble. This is confirmed by their analysis of the relevant non-equilibrium transport equations, which shows that the baryon asymmetry behaves completely regularly at the speed of sound.

However, these results crucially depend on the \emph{Ansatz} for the shape of the non-equilibrium particle distribution function. In this recent study~\cite{Cline:2020jre} the working hypothesis was (based on previous baryogenesis analyses~\cite{Fromme:2006wx}) that these functions are fully characterized by two perturbations only: a chemical potential and a velocity parameter encapsulating how the particle's velocity deviate locally from the plasma velocity $u^\mu$. All other possible perturbations are assumed to be linearly related to these velocity perturbations.

Apart from being a restrictive and largely unjustified assumption, this factorization hypothesis does not sufficiently specify the shape of the distribution functions to allow for a computation of the relevant collision terms. For these, one has to assume a specific \emph{Ansatz}, e.g.~the fluid \emph{Ansatz} of the form
\begin{equation}
	f(x,p) = \frac{1}{e^{\beta (p^\mu u_\mu + \delta)} \pm 1},
	\label{eq:flow}
\end{equation}
and expand the perturbations $\delta$ in powers of momenta, truncating the expansion at some appropriate order~\cite{Moore:1995si, Arnold:2000dr}. Thus, the working hypothesis of reference~\cite{Cline:2020jre} is actually inappropriate for practical purposes, since it forces one to use different \emph{Ans\"atze} for different terms of the Boltzmann equation. 

When the transport analysis is performed thoroughly and consistently with this \emph{Ansatz}, keeping only terms up to first order in momenta (a so-called \emph{fluid Ansatz} which amounts to including a chemical potential, a velocity perturbation and a temperature fluctuation as well), the speed of sound emerges as a singularity of the Liouville operator, corresponding to the kinetic term in the Boltzmann equations. This implies that, for supersonic walls, all fluctuations trail the source~\cite{Konstandin:2014zta}: there are no non-equilibrium dynamics in front of the wall, and the baryon asymmetry should therefore vanish. This is in sharp contrast to recent findings in \cite{Cline:2020jre}, but still does not constitute a definitive position on the issue, because one must still ask whether this first order truncation is justified at all, and how the system would change if higher order terms were included.

The goal of the present paper is to extend this analysis of the transport equations using an \emph{extended fluid Ansatz}, including more perturbations appearing at higher orders in momenta. To keep consistency, collision terms from annihilation and scattering processes associated to these new fluctuations are computed numerically and analytically. Since the issue under investigation is encapsulated in the behavior of the kinetic term, we choose to simplify the analysis by neglecting complications arising from CP-violation as well as neglecting collision terms that would be relevant to a full baryogenesis study, such as chirality flips, $W$-boson interactions and strong sphalerons.

Our treatment is rigorously adequate for the study of friction in front of the wall, a problem which has been tackled in a fully relativistic approach both with the first-order fluid approximation~\cite{Konstandin:2014zta} and in the formalism adopted in \cite{Cline:2020jre}. But baryogenesis and wall friction are very similar phenomena, since both rely on out-of-equilibrium distributions of the particles in the plasma close to the propagating wall, and in both cases the source that drives the distribution functions away from equilibrium is the interaction between the particles and the Higgs. The main difference is that in baryogenesis one is mostly interested in the CP-violating component of the deviation from equilibrium, but for the purposes of discussing the behavior of the solutions around the speed of sound this is not relevant. Interestingly, our findings support the results in \cite{Cline:2020jre} that the fluctuations behave continuously across the sound speed, albeit for different reasons. 

The paper is organized as follows. In section~\ref{sec:Flow} we review the standard fluid approximation for the non-equilibrium distribution function~\cite{Moore:1995si,Konstandin:2014zta}, and how the Boltzmann equation can be used to determine the three fluctuations in chemical potential, local particle velocity and local temperature. We will see how the speed of sound emerges naturally in this setup as the singularity of the kinetic term. In section~\ref{sec:Boltzmann} we present some arguments leaning towards a criticism of this simple approximation and motivate a generalization with more fluctuations. We also present here some specific criticisms on the standard approach for transport baryogenesis found in the literature~\cite{Bodeker:2004ws, Fromme:2006wx, Cline:2020jre}, arguing that one cannot escape from making a specific Ansatz for the shape of the distribution functions. In section~\ref{sec:Model} we present our generalized fluid Ansatz, showing explicit expressions for the kinetic matrix, the collision terms and the source up to six fluctuations. 
Our results are presented in section~\ref{sec:results}, and we conclude in section~\ref{sec:conclusions}.
An argument from hydrodynamics for the relevance of the speed of sound is presented in Appendix~\ref{sec:hydro}. 
General expressions for the collision terms to leading-log accuracy up to arbitrary orders in the generalized Ansatz can be found in Appendix~\ref{sec:coll}.

\section{The fluid approximation}
\label{sec:Flow}

Ultimately, we want to study the Boltzmann equation, that in its relativistic form reads
\be
p^\mu \partial_\mu \, f_i(x^\mu, p^\mu) + 
m \, F^\mu \partial_{p^\mu} \, f_i(x^\mu, p^\mu)  = {\cal C}[f_j] \, ,
\ee
where $p^\mu$ is the four-momentum of the particles (evaluated on-shell $p^0 = E = \sqrt{\vec p^{\, 2} + m^2}$ ), $f$ is the particle distribution function, $F$ denotes the forces in the system and ${\cal C}$ is the collision term. 

The Boltzmann equation is a non-linear partial differential equation (due to the collision term), and in order to make progress different approximations are typically used. The first approximation is to assume that the system is close to equilibrium. The equilibrium distribution is given in terms of the four-velocity of the fluid $u^\mu$ (relative to the global restframe of the plasma) and the temperature $T$. In the current setup, the forces will drive the system out-of-equilibrium while the collision terms will relax the system to some local equilibrium that in principle can be different on both sides of the wall (in fact it has to be so, as demanded by energy momentum conservation, see Appendix~\ref{sec:hydro}). In the analysis of baryogenesis, this problem can be eliminated by introducing a background fluid (made up mostly from the gluons and quarks that are not much affected by the Higgs) that will reflect this change as demonstrated in \cite{Moore:1995si}.

Since we assume the system close to equilibrium, one can split the distribution function into a deviation and an equilibrium value 
\be
f(p^\mu, x^\mu) =  \delta f(p^\mu, x^\mu) + f^{eq}(p^\mu u_\mu/T) \, .
\ee
After this one arrives at a system of equations that reads
\be
p^\mu \partial_\mu \, \delta f_i(x^\mu, p^\mu)  = {\cal C}[\delta f_j] + {\cal S}[f^{eq}_j] \, ,
\label{eq:Boltzmann}
\ee
where the source term ${\cal S}$ contains all the forces and depends at leading order only on some forces acting on the equilibrium
distributions.

Here we want to note that any peculiarities at the speed of sound are due to a vanishing (or very small) eigenvalue in the first term of this equation (the so-called Liouville term). In particular, it has nothing to do with the specific form of the sources or collision terms. This is also why the relativistic analysis for the wall friction can be carried over to baryogenesis without much effort.

In the following we summarize the steps in ref.~\cite{Konstandin:2014zta} to present the analysis for the fluid approximation in Lorentz covariant form.
The fluid approximation assumes the different species have individual temperatures and fluid velocities and also chemical potentials. Assuming that the fluctuations are small relative to the background, one obtains
\be
\delta f \simeq -\left( \mu/T + p^\mu \delta u_\mu/T  - p^\mu u_\mu \delta T/T^2 \right) f_{eq}'(p^\mu u_\mu/T) \, .
\label{eq:df}
\ee
The fluctuations $\mu$, $\delta T$ and $\delta u^\mu$ describe (for every species) the deviation from equilibrium. Note that $\delta u^\mu$ obeys $u_\mu \delta u^\mu = 0$ so that, in the planar wall case we will consider, it constitutes only one degree of freedom. Therefore in this approximation the system is described by three fluctuations in total.

Before we come to the usual approach of solving these equations, let us discuss whether this Ansatz makes sense. 
The main argument here is that (depending on the system under consideration), there often is a hierarchy between different classes of interactions. The strongest interactions are often gauge scatterings that do not change the particle content of the plasma. In the Standard Model, these are e.g. 2-by-2 scatterings mediated by the gluons. Notice that the $t$-channel interactions even feature a logarithmic enhancement on top of the rather large gauge coupling in the strong sector~\cite{Arnold:2000dr, Arnold:2003zc}. When inspecting the collision terms, these interactions will equilibrate the quarks among themselves. Very quickly, all quarks should have a distribution function close to the fluid approximation and share the same temperature and velocity but not chemical potential, since this requires interactions that change particle number. 

This leads to the conventional wisdom that kinetic equilibrium is attained faster than chemical equilibrium.
(Notice that there is a subtle difference in the definition of the chemical potentials of particles and anti-particles when it comes to the study of the friction compared to baryogenesis, see section \ref{sec:Model}).
 Still, ultimately, there is no proof that the fluid approximation is holding in the baryogenesis and/or friction setup and that differences between the fluctuations in the temperatures and flow velocities should equilibrate slower than any other fluctuations. This leaves the question of how important these fluctuations are for a correct determination of the friction or baryon asymmetry. This is an open question that was already raised in the seminal work by Moore and Prokopec~\cite{Moore:1995si}, and is discussed to some extent in their Appendix B.     
 
Even though one cannot make a strong argument in favor of the fluid approximation with only three perturbations, let us briefly discuss the standard procedure in order to understand the generalization which will follow, as well as to understand where the strong dependence on the wall velocity in the setup comes from. Since the fluid approximation has three degrees of freedom per species, one can take three moments of the Boltzmann equations to obtain a (non-partial) system of differential equations. The most natural choice is to simply multiply by four-momenta and integrate, so that the Boltzmann equation~(\ref{eq:Boltzmann}) reads 
\bea 
\partial_\mu \int \frac{d^3p}{E} p^\mu \delta f(p,x) &=& \text{collisions + source} \, ,\label{eq:J} \\
\partial_\mu \int \frac{d^3p}{E} p^\mu p^\nu \delta f(p,x) &=& \text{collisions + source} \, .\label{eq:T}
\eea
Note that the denominator includes only one power of $E$ from the Lorentz invariant integration measure. At this point our treatment differs from \cite{Cline:2020jre} (and also from~\cite{Bodeker:2004ws, Fromme:2006wx}) who advocate different moments in their analyses, obtained by multiplying the Boltzmann equations with extra factors of $1/E$ as well. Our choice of moments seems more natural, since the resulting equations can be easily interpreted as the divergence of the charge current and the energy-momentum tensor. As we will see, this is one important reason why the speed of sound is of relevance in the analysis of \cite{Moore:1995si}, while it is 
not in \cite{Cline:2020jre}. Obviously, the analysis presented here does not rely on small wall velocities.

We can then linearize the system using the fluctuations in (\ref{eq:df}) and focus on the case of an approximately planar wall, which is justified after the bubble reaches a steady-state, and also because we are interested in the behaviour of fluctuations close to the wall where its curvature becomes negligible. This means that there is only one direction $\bar u^\mu$ perpendicular to the fluid velocity, i.e.~with $\bar u^\mu u_\mu = 0$ and $\bar u^\mu \bar u_\mu = -1$. In particular the velocity fluctuations can be written as $\delta u^\mu = \bar u^\mu \delta v$. Moreover in the steady-state situation the fluctuations depend only on $\xi = x^\mu v_\mu$, where $v_\mu=\gamma (v_w u_\mu - \bar u_\mu)$ characterizes the four-velocity of the wall, which then leads to $\partial_\mu = v_\mu \partial_\xi$. Finally, we note that equation~(\ref{eq:T}) contains two components, which we can project onto $u_\nu$ and $\bar u_\nu$, leading to a total of three equations for three fluctuations. Putting everything together, we see that the Liouville part of the Boltzmann equations, corresponding to the kinetic component of the flow, involves derivatives $\partial_\xi (...)$ of the fluctuations multiplied by coefficients of the form 
\be
	\int \frac{d^3 p}{E} (p^\mu u_\mu)^m (p^\nu \bar u_\nu)^n (-f^\prime_{eq})
	 = 4\pi T^{m+n+2} \left\{ \begin{array}{ll} \dfrac{c_{m+n+1}}{n+1},  &n\text{ even}\\
	 											0, & n\text{ odd},\end{array}\right.
	\label{eq:cint}
\ee
where, in the massless case and for $n\geq 2$,
\bea
c_n^b \equiv \frac{1}{T^{n+1}}\int dp \, p^n  f^\text{BE}_p (1+ f^\text{BE}_p) &=& n!\,\zeta_n\,,\\
c_n^f \equiv \frac{1}{T^{n+1}}\int dp \, p^n  f^\text{FD}_p (1- f^\text{FD}_p) &=& \left(1-\frac{1}{2^{n-1}}\right) n!\,\zeta_n\,,
\label{eq:p_ints}
\eea
with $f^\text{BE}$ and $f^\text{FD}$ the Bose-Einstein and Fermi-Dirac distribution functions, respectively. For the collision terms associated to annihilations, it will also be convenient to define
\be
\widetilde{c}_{n+1} \equiv \frac{1}{T^{n+1}}\int dp \, p^n  f^\text{FD}_p (1+ f^\text{BE}_p) = \left(1-\frac{1}{2^{n+1}}\right) n!\,\zeta_{n+1}\,,
\label{eq:ctilde}
\ee
which we quote here for reference.

In Appendix~\ref{sec:coll}, the collision terms are derived to linear order, so this momentum decomposition of the Boltzmann equation leads to a system of the form
\be
A \cdot \frac{d}{d\xi}q + \Gamma \cdot q = S \, ,
\label{eq:system}
\ee
with the vector $q=(\delta \mu/T, -\delta T/T, \delta v)$ describing the fluctuations of the system. The matrix $\Gamma$ stems from the linearized collision term while $S$ contains the sources. On the other hand, the matrix $A$ comes from the Liouville term and has the form (for a relativistic plasma in equilibrium) 
\be
\begin{pmatrix}
v_w c_2 & v_w c_3 & c_3/3 \\
v_w c_3 & v_w c_4 & c_4/3 \\
c_3/3 & c_4/3 & v_w c_4/3 \\
\end{pmatrix}.
\label{eq:kin_o1}
\ee
Now, the authors of \cite{Moore:1995si} made the interesting observation that for $v_w = 1/\sqrt{3}$ the determinant of this matrix vanishes and one eigenvalue changes sign. 
The corresponding eigenvector corresponds to $v_w \delta v = -\delta T/T$.
For supersonic wall velocities all eigenvalues have the same sign such that all fluctuations trail the wall. Hence the only contribution to baryogenesis comes from a local source in the wall and the resulting asymmetry is strongly suppressed. Notice that this discussion does not rely on the special form of the collision term or the source. Only the analysis of the Liouville term in the Boltzmann equation is relevant to make this argument. Moreover, the effect should occur in every species individually and does not arise from the interplay of different species. 

This fits rather nicely with the observations of how the system behaves on the largest scales, as inferred from hydrodynamics (see Appendix~\ref{sec:hydro}). 
In particular, the behavior of the system changes qualitatively once the wall velocity surpasses the speed of sound.

\section{Why higher moments are essential}
\label{sec:Boltzmann}

Since the validity of the fluid Ansatz with only three momentum-independent fluctuations is somewhat controversial, let us now venture towards a generalisation of the previous argument. As a consequence of the discussion in this section, we will propose a generalised fluid Ansatz with higher order perturbations, analogous to including a momentum dependence of the fluctuations in the previous section. 

Let us start by noticing that sound waves are a collective phenomenon that obviously needs at least some notion of equilibrium. Assuming that the plasma has a rest frame where the distribution functions only depend on the energy, in a general frame the distribution function is of the form $f(p^\mu u_\mu/T)$ where $u^\mu$ is the fluid four-velocity and the temperature $T$ is introduced for dimensional reasons. Otherwise, the function $f$ is arbitrary.

Now consider two fluctuations that mimic the temperature and velocity fluctuations from the last section,
\be
\delta f = \left(p^\mu \bar u_\mu \delta v/T  - p^\mu u_\mu \delta T/T^2 \right) f'(p^\mu u_\mu/T) \, ,
\label{eq:zeromode}
\ee
and let us again focus on the steady-state situation, with $\partial_\mu = \gamma (v_w u_\mu - \bar u_\mu)\partial_\xi$. The Liouville term then contains three different Lorentz structures,
\bea
& p^\nu{\bar u_\nu} \, p^\mu u_\mu (v_w \partial_\xi \delta v + \partial_\xi \delta T/ T ) f'(p^\mu u_\mu/T) /T, & \nn \\
& p^\nu{\bar u_\nu} \, p^\mu {\bar u_\mu} (\partial_\xi \delta v) f'(p^\mu u_\mu/T) /T, & \nn \\
& p^\nu{u_\nu} \, p^\mu u_\mu (v_w\partial_\xi \delta T/ T ) f'(p^\mu u_\mu/T) /T. & \nn 
\eea
As in the fluid approximation, for $v_w \delta v = - \delta T/ T$ the first term vanishes but the others do apparently not.
The remaining terms combine into 
\be
p^\nu p^\mu ( {\bar u_\nu} \,{\bar u_\mu} - v_w^2\, u_\mu u_\nu ) (\partial_\xi \delta v) f'(p^\mu u_\mu/T) /T.
\ee
However, for a relativistic plasma, in the plasma frame $p^\mu u_\mu = E$ and $p^\mu \bar u_\mu = p_z$. Moreover, for massless particles $\left<E^n\, p_z^2 f^\prime \right> = \left<E^{n+2} f^\prime \right> c_s^2$ and odd moments of $p_z$ vanish (this follows e.g.~from the fact that the energy momentum should be traceless). This means that the lowest moment that can potentially spoil the vanishing eigenvalue in the kinetic term contains four factors of $p^\mu \bar u_\mu$. This is one of the motivations to generalize the fluid approximation, to introduce a larger class of fluctuations and to use higher moments of the Boltzmann hierarchy to study the system more rigorously.

Before jumping into this analysis, let us briefly comment on the alternative approach found in the literature, which claims to perform the analysis without an explicit Ansatz for the perturbations. One can quite generally write the non-equilibrium distribution function as
\be
f = \frac{1}{\exp(\beta (p^\mu u_\mu - \mu)) \pm 1} + \delta f,
\ee
letting $\delta f$ encode all fluctuations away from equilibrium except for the momentum-independent part of the chemical potential, which is encapsulated by $\mu$. In standard transport baryogenesis treatments~\cite{Bodeker:2004ws, Fromme:2006wx, Cline:2020jre} one then take moments by multiplying the Boltzmann equation by powers of $p_z/E$ in the plasma frame, resulting in fluctuations defined as\footnote{Note the additional factors of $1/E$ for $\ell \geq 2$, which are absent in our approach delineated in the previous section.}
\be
u_\ell \equiv \left< \left(\frac{p^\mu \bar u_\mu}{E}\right)^\ell \delta f \right> \, .
\label{eq:CKAnsatz}
\ee
Of course, this description is so general that it includes the fluid approximation as a particular case. But, without imposing any further restriction on the form of $\delta f$, one cannot make any progress towards determining the shape of $\delta f$. So some Ansatz is unavoidable. This difficulty is then typically sidestepped by imposing a factorization condition which essentially states that all fluctuations $u_\ell$ for $\ell\geq 2$ are proportional to $u_1$. This entire approach is problematic for a few reasons.

First, the choice of momenta is very peculiar. While $u_1$ is the $z$-component of the four-current in the wall frame, $u_2$ is not simply related to the relevant conserved quantities like the energy momentum tensor, and consequently the physical meaning of the resulting equations in this formalism is less transparent. Moreover, using the energy-momentum tensor conservation as a moment, as in equation~(\ref{eq:T}), has the advantage that the corresponding collision terms vanish (not for individual species but collectively). 

Next, this approach is supposedly the most general since it does not rely on an Ansatz. But in fact the specific choice of factorisation, relating all higher-order fluctuations to one velocity perturbation $u_1$,   
is equivalent to an \emph{Ansatz}, with the disadvantage that its physical significance is not very transparent. 
Besides, this Ansatz is not specific enough to allow for the computation of the collision terms. For that, one needs to know the explicit form of the non-equilibrium distribution, such as we do have in the fluid approximation. Even though this seems unrelated to our issue at hand, this makes the method impractical in actual calculation. In previous baryogenesis analyses~\cite{Bodeker:2004ws, Fromme:2006wx, Cline:2020jre} the collision terms have been ultimately derived assuming the fluid approximation. 

Finally, as discussed in the last section, for the peculiar fluctuations~(\ref{eq:zeromode}) the 
kinetic term vanishes up to fourth order in $p_z$ at the speed of sound, so by construction $u_1 = u_2 = 0$ and the system is oblivious to the eigenvector with zero eigenvalue. 
In other words, the functions $u_1$ and $u_2$ do not encode this eigenvector and only higher moments would be sensitive to it.  
  
In summary, we cannot escape from using a specific Ansatz for the non-equilibrium distributions, yet there is little justification in keeping only three momentum independent fluctuations in the fluid approximation. In the following, we will generalize this fluid Ansatz to include terms at higher order in momenta. We will see that the results for baryogenesis across the speed of sound are modified even at a qualitative level.

\section{Generalized fluid Ansatz}
\label{sec:Model}

A general non-equilibrium distribution can always be written in the form shown in equation~(\ref{eq:flow}). But, in much the same way as argued above, this is too generic to allow for the computation of the resulting perturbations from the Boltzmann equation. Some progress can be made by expanding the perturbations $\delta$ in powers of momenta. The standard fluid approximation is then obtained by truncating this expansion at first order, as in~(\ref{eq:df}). 
Here, we would like to extend the set of fluctuations and study the behaviour of the solutions. This is not the first time that higher moments have been considered in treatments of the Boltzmann equation, see e.g.~\cite{Grad:1949, DeGroot:1980dk, Moore:1995si} for an interpretation of the higher-order equations in terms of entropy production due to dissipative work and heat flow. However, the present work is the first that uses this extended approach in the context of baryogenesis.

To be specific, we parameterise the fluctuations as
\bea
\delta f &=& \left( w^{(0)} + p^\mu w^{(1)}_\mu + p^\mu p^\nu w^{(2)}_{\mu\nu}  + \cdots \right) \nn \\
&& \quad \times \quad  f_{eq}'(p^\mu u_\mu/T) \, .
\label{eq:Ansatz}
\eea
This generalizes the fluid approximation, and due to the symmetries only $n+1$ degrees are relevant in 
$w^{(n)}$. For instance, expanding up to $n=2$ will lead to 6 fluctuations in total.
In order to truncate the Boltzmann hierarchy we then multiply the Boltzmann equations by the factors 
$(p^\mu u_\mu)^a(p^\nu \bar u_\nu)^b$ with $a,b\geq 0$ and $(a+b)\leq n$ and take the corresponding moments to arrive at
a linear equation system for the fluctuations.

The result is again a system of the form shown in equation~(\ref{eq:system}), but now with enlarged $(n+1) (n+2)/2$ square matrices,
whose entries will involve the coefficients defined in~(\ref{eq:p_ints}) and (\ref{eq:ctilde}). 
Indeed, for $n=2$ the kinetic matrix has the form
\be
	A = \begin{pmatrix}
		v_w c_2 & v_w c_3 & c_3/3 & v_w c_4 & c_4/3 & v_w c_4/3\\
		v_w c_3 & v_w c_4 & c_4/3 & v_w c_5 & c_5/3 & v_w c_5/3\\
		c_3/3 & c_4/3 & v_w c_4/3 & c_5/3 & v_wc_5/3 & c_5/5\\
		v_w c_4 & v_w c_5 & c_5/3 & v_w c_6 & c_6/3 & v_w c_6/3 \\
		c_4/3 & c_5/3 & v_w c_5/3 & c_6/3 & v_w c_6/3 & c_6/5 \\
		v_w c_4/3 & v_w c_5/3 & c_5/5 & v_w c_6/3 & c_6/5 & v_w c_6/5
	\end{pmatrix}.
\ee

The source term is defined in equation~(\ref{eq:Boltzmann}) as
\be
{\cal S}[f^{eq}]\equiv -mF^\mu \partial_{p^\mu} f_{eq} - p^\mu \partial_\mu\,f_{eq} = -f^\prime_{eq}\, m\,u^\mu F_\mu,
\ee
and the force can be divided into a CP-even and a CP-odd term as $m F_\mu = \partial_\mu m^2/2 + m F_\mu^{\cancel{CP}}$. As mentioned before, baryogenesis relies on C and CP violation, so the CP-odd component of the source is essential in this calculation. However, in this work we are interested in studying the behaviour of the solutions across the speed of sound, which depends exclusively on the kinetic term. In this sense the calculation of the baryon asymmetry and of friction are similar, and we will therefore consider here a CP-even source for simplicity. Then, after multiplying the Boltzmann equation by factors of $p^\mu u_\mu$ and $p^\mu \bar u_\mu$ and integrating, we again get integrals as in equation~(\ref{eq:cint}), resulting in
\be
	\mathcal{S} = \frac{u^\mu \partial_\mu m^2}{2}
		\begin{pmatrix}
			c_1 & c_2 & 0 & c_3 & 0 & c_3/3		
		\end{pmatrix}^T,
\ee
for six perturbations. Here $c_1^f = \log 2$ and $c_1^b = \log(2T/m)$, whereas $c_2$ and $c_3$ are given in equations~(\ref{eq:p_ints}).

Focusing on the CP-even terms has an extra advantage in simplification. For baryogenesis one would have to compute the CP-odd components of the chemical potential and of the other 
fluctuations. In particular, the chemical potential in baryogenesis calculations is a 
proper chemical potential, while the fluctuation $\mu$ in friction calculations parametrizes 
a tilt in the distribution function that is equal for particles and anti-particles. Hence, in the collision terms the fluctuation $\mu$ in friction calculation will be damped by annihilation
processes. In baryogenesis, on the other hand, damping of the chemical potential obviously requires particle number changing interactions that are far weaker and less abundant. This is because, due to the CP
 violating source of baryogenesis, the chemical potentials have opposite signs for different chiralities
 and are reduced by the mass terms of the fermions in the broken phase. Moreover, the strong sphaleron can flip the chirality of the quarks and erase their chemical potentials. Therefore, in focusing on the CP-even part of the fluctuations only, we neglect these complications and simplify the analysis of the collision terms as well.
 
In summary, we consider collision terms from top-quark annihilations and scatterings off other quarks and off gluons. The collision matrix for annihilations is
\bgroup
\def\arraystretch{1.35}
\be
	\Gamma_\text{ann} = \frac{16\alpha_s^2\,T}{9\pi}
		\begin{pmatrix}
			2\gamma_\text{ann}^{0000} & 2\gamma_\text{ann}^{1000} & 0 & 2\gamma_\text{ann}^{2000} & 0 & 2\gamma_\text{ann}^{0020}\\
			2\gamma_\text{ann}^{1000} & \gamma_\text{ann}^{2000}+\gamma_\text{ann}^{1100} & 0 & \gamma_\text{ann}^{3000}+\gamma_\text{ann}^{2100} & 0 & \gamma_\text{ann}^{1020}+\gamma_\text{ann}^{1002}\\
			0 & 0 & \gamma_\text{ann}^{0020} & 0 & \gamma_\text{ann}^{1020} & 0\\
			2\gamma_\text{ann}^{2000} & \gamma_\text{ann}^{3000}+\gamma_\text{ann}^{2100} & 0 & \gamma_\text{ann}^{4000}+\gamma_\text{ann}^{2200} & 0 & \gamma_\text{ann}^{2020}+\gamma_\text{ann}^{2002}\\
			0 & 0 & \gamma_\text{ann}^{1020} & 0 & \gamma_\text{ann}^{2020} & 0 \\
			2\gamma_\text{ann}^{0020} & \gamma_\text{ann}^{1020}+\gamma_\text{ann}^{0120} & 0 & \gamma_\text{ann}^{2020}+\gamma_\text{ann}^{0220} & 0 & \gamma_\text{ann}^{0040}+\gamma_\text{ann}^{0022}
		\end{pmatrix}
	\label{eq:Gann}
\ee
and for scatterings
\be
	\Gamma_\text{scatt}=\frac{16\alpha_s^2\,T}{9\pi}\frac{9A}{4}
		\begin{pmatrix}
			0 & 0 & 0 & 0 & 0 & 0\\
			0 & \gamma_\text{scatt}^{1100} & 0 & \gamma_\text{scatt}^{2100} & 0 & \gamma_\text{scatt}^{0120}\\
			0 & 0 & \gamma_\text{scatt}^{0011} & 0 & \gamma_\text{scatt}^{1011} & 0 \\
			0 & \gamma_\text{scatt}^{2100} & 0 & \gamma_\text{scatt}^{2200} & 0 & \gamma_\text{scatt}^{0220}\\
			0 & 0 & \gamma_\text{scatt}^{1011} & 0 & \gamma_\text{scatt}^{1111} & 0\\
			0 & \gamma_\text{scatt}^{0120} & 0 & \gamma_\text{scatt}^{0220} & 0 & \gamma_\text{scatt}^{0022}
		\end{pmatrix},
	\label{eq:Gsct}
\ee
\egroup
where $A=1$ for scatterings by gluons and $A=5/3$ for scatterings by quarks. Analytic expressions for the coefficients $\gamma_\text{ann}$ and $\gamma_\text{scatt}$ at leading-log can be found in equations~(\ref{eq:gann}) and (\ref{eq:gscat}) in Appendix~\ref{sec:coll}.

Now, a further complication arises when we attempt to include the sphalerons in this generalized framework. 
In standard baryogenesis calculations, based on two fluctuations only, the sphalerons couple only 
to the zeroth-order fluctuation, i.e.~that which is not multiplied by any power of momenta, and which is interpreted as the chemical potential of the corresponding field. 
In principle, in our framework one would have to recalculate how the effective interaction of the strong sphaleron damps the fluctuations in our generalized Ansatz (\ref{eq:Ansatz}). This task is beyond the scope 
of what we want to achieve here and we will mimic the true collision terms of the sphalerons in different ways.

The strong sphaleron will be mimicked using the collision term from the friction calculation. This 
will lead to a damping of the chemical potential similar to what the strong sphaleron achieves in baryogenesis. 
At the same time it is a set of physical collision terms that just occur in a different setting. So, strictly speaking, the convergence properties that we will observe are the ones of the friction network and not the 
ones of baryogenesis. In order to facilitate transport, we choose the corresponding gauge couplings rather small, such that the charges can travel into the symmetric phase as in the case of non-local baryogenesis.

To mimic the weak sphaleron, we use the standard result for the sphaleron rate,
\be
\Gamma_{ws} \simeq 10^{-6} T \exp(-a \phi(z)/T) \, ,
\label{eq:Gws}
\ee
where $a\approx 37$ and $\phi(z) = \frac{\phi_0}{2}(1-\tanh\frac{z}{L_w})$ is the bubble profile with wall thickness $L_w$ and vev $\phi_0$ at the critical (or nucleation) temperature. We couple the sphaleron to $J^\mu u_\mu$, where $J^\mu = \left< p^\mu/E\right>$ is the current of the plasma, {since this is after all the moment associated to particle number density and should therefore be interpreted as the full chemical potential~\cite{DeGroot:1980dk}\footnote{The interpretation of the zeroth-order perturbation $w^{(0)}$ as the chemical potential is only valid under the condition that\cite{DeGroot:1980dk} $\int \frac{d^3 p}{E} p^\mu u_\mu \delta f = 0$, which is not automatically satisfied in our Ansatz. The coincidence is only granted in a two-fluid approximation, which has often been employed in previous baryogenesis studies.}}. Coupling the weak sphaleron to the zeroth-order fluctuation would be rather arbitrary and would make the result highly dependent on the precise basis of fluctuations. In any case, the coupling to the weak sphaleron is of course
not relevant to study convergence of the out-of-equilibrium distribution functions but rather to reproduce the qualitative behavior of baryogenesis
for a low number of fluctuations and supersonic wall velocities where the resulting BAU is suppressed.

\section{Results}
\label{sec:results}

With this setup, we have all we need to solve the resulting Boltzmann system and find the fluctuations and the toy baryon asymmetry. The latter is obtained by solving
\be
\partial_z n_B = \frac3{2v_w} \Gamma_{ws} ( \kappa \, u_\mu J^\mu - {\cal A} \, n_B )   \, ,
\ee
with ${\cal A} = 15/2$, and $\kappa=3/(4\pi c_2)$ is a factor chosen so that the result will agree with the traditional coupling to the chemical potential at zeroth-order. 
In order to find the current $J^\mu$ we solve the system
\be
	\frac{d}{d\xi}q + (A^{-1}\cdot \Gamma)\,q = A^{-1}\cdot S \, .
\ee
If $\lambda_i, \chi_i$ are respectively the eigenvalues and eigenvectors of $A^{-1}\cdot \Gamma$, the Green's function is
\be
	G(z) = \left\{ \begin{array}{cc}
			\sum_{\lambda_i>0} \alpha_i\,\chi_i \exp({-\lambda_i\,z}), & z>0\\
			\sum_{\lambda_i<0} \alpha_i\,\chi_i \exp({-\lambda_i\,z}), & z<0 \, ,
				\end{array} \right.
\ee
and the fluctuations are
\be\begin{split}
	q(z) &= \int_{z}^{\infty} dz^\prime\,\sum_{\lambda_i>0}\,
				(\chi_i^{-1}\cdot A^{-1}\cdot S)(z^\prime)\,\chi_i\,
				\exp\left[-\lambda_i\,(z^\prime-z)\right]\\
			&~~~ -\int_{-\infty}^{z} dz^\prime\,\sum_{\lambda_i<0}\,
				(\chi_i^{-1}\cdot A^{-1}\cdot S)(z^\prime)\,\chi_i\,
				\exp\left[-\lambda_i\,(z^\prime-z)\right].
\end{split}\ee
We see that for positive eigenvalues the fluctuations at $z$ get contributions from the source at $z^\prime > z$, and we say that the fluctuations trail the source. The opposite occurs for negative $\lambda_i$, in which case the fluctuations are ahead of the source. Because the baryon number is obtained from a convolution of the fluctuations with the sphaleron rate $\Gamma_{ws}$, and because the latter is only active in front of the wall, as can be seen from equation~(\ref{eq:Gws}), the trailing fluctuations produce a negligible baryon asymmetry.

Thus, when all eigenvalues are positive, the resulting asymmetry should vanish. This is exactly what happens for supersonic wall velocities in the fluid approximation with three perturbations, which leads to the interpretation of the speed of sound as an upper limit for viable transport baryogenesis. But when we add additional fluctuations, not all eigenvalues become positive beyond this threshold. Some eigenvalues may remain negative and yield a contribution to the BAU, albeit smaller than for lower velocities because some eigenvalue  sign flipping does invariably take place.

\begin{figure}
\begin{center}
	\includegraphics[scale=.38]{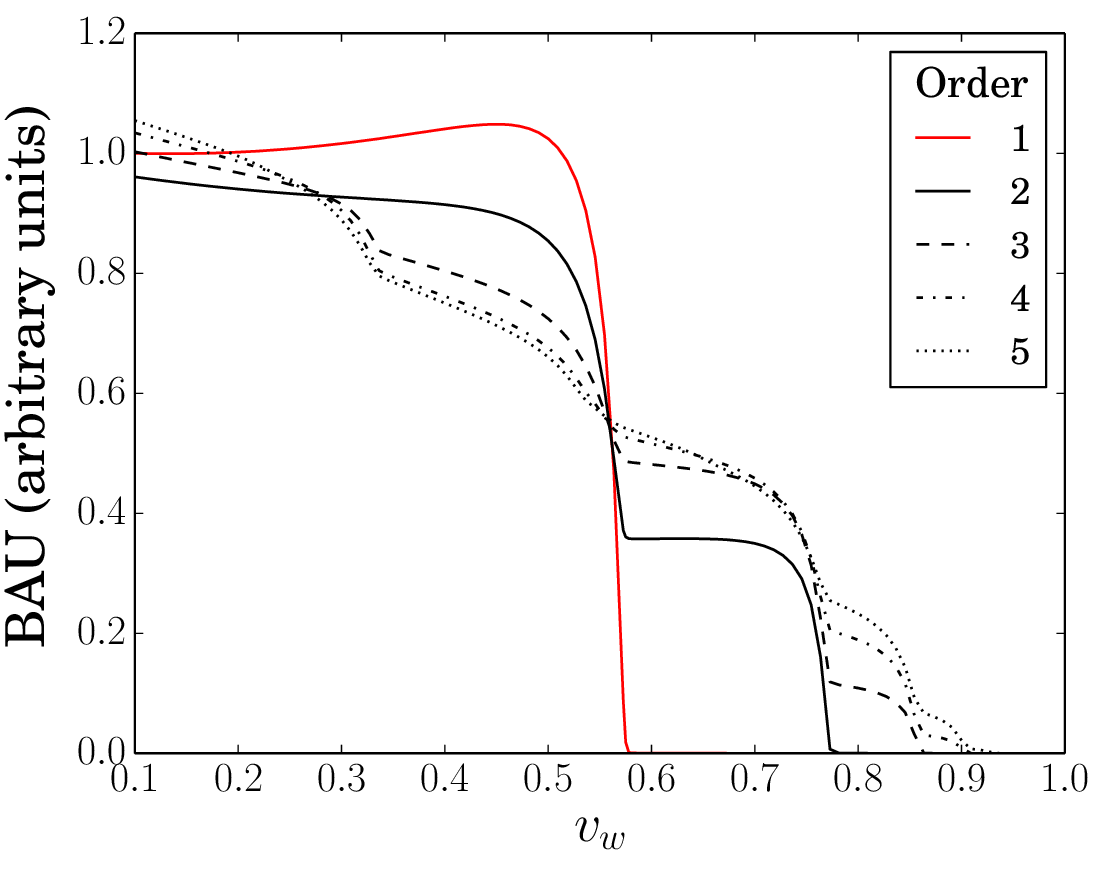}
	\includegraphics[trim=37 0 0 0, clip, scale=.38]{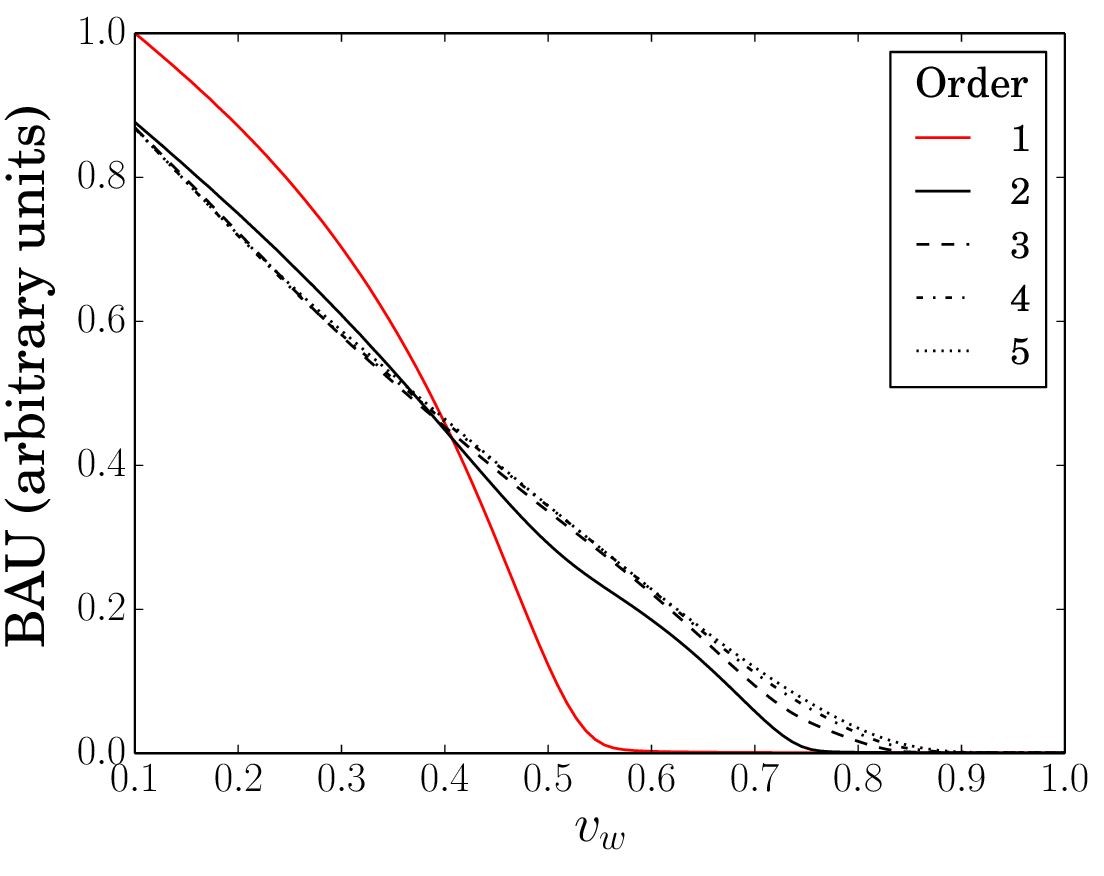}
	\caption{Baryon asymmetry for $\alpha_s = 0.01$ (left) and $\alpha_s=0.06$ (right) as a function
	of the wall velocity. The different lines correspond to a different basis of fluctuations and 
	different numbers of moments.}
	\label{fig:results}
\end{center}
\end{figure}

This is illustrated in figure~\ref{fig:results}, with the resulting toy baryon asymmetry for two values of the relevant coupling entering the collision terms (in this case, the strong coupling $\alpha_s$). The red line represents the case of three fluctuations, where it is clear that there is no asymmetry beyond the speed of sound. However, once we add more perturbations, this picture changes and a resulting asymmetry does become possible. There also appear other thresholds, related to new singularities of the larger kinetic matrix. Curiously, as one adds more and more perturbations, the sharp drops in these threshold values seem to smoothen out and one approaches a continuous curve, similar to the result obtained in~\cite{Cline:2020jre}.

We highlight, however, that the reasoning for this similar behaviour is fundamentally different. Here the smoothness is an emergent asymptotic behaviour obtained from a well-defined expansion in momenta, rather than from an \emph{ad hoc} factorization assumption. Furthermore, the speed of sound does not constitute a sharp discontinuity, but it remains a feature of the system, even if hidden in the first few momenta only.  Our approach is also thoroughly consistent, in the sense that we use the same Ansatz to compute all the terms in the Boltzmann equation, including the collision terms.  

It is also interesting to point out that the result clearly converges as we add more fluctuations. As can be seen by comparing the two plots in figure~\ref{fig:results}, the parameter determining this convergence is the inverse coupling appearing in the collision terms. This is not unexpected: the stronger the interactions are, the quicker and more effective the thermalisation processes will be, so we would effectively need less perturbations to describe the distribution functions well. 

\begin{figure}
\begin{center}
	\includegraphics[scale=.38]{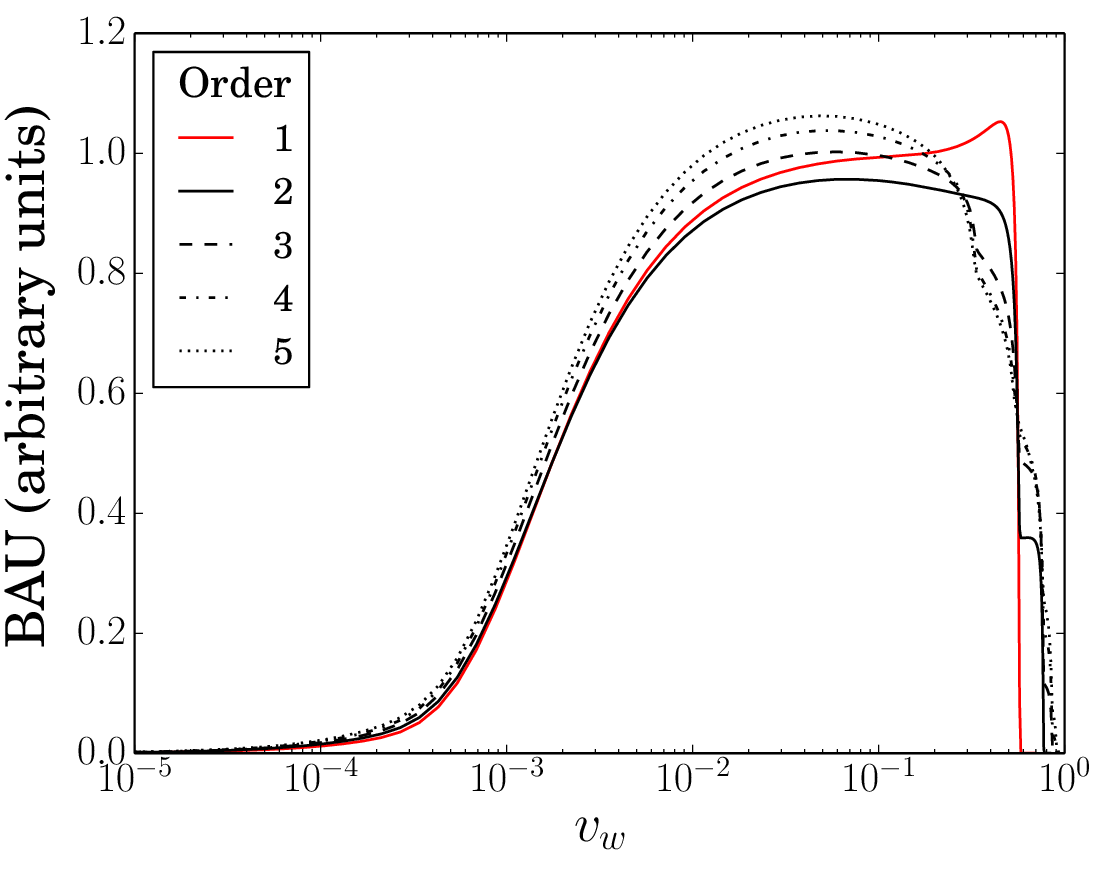}
	\includegraphics[trim=37 0 0 0, clip, scale=.38]{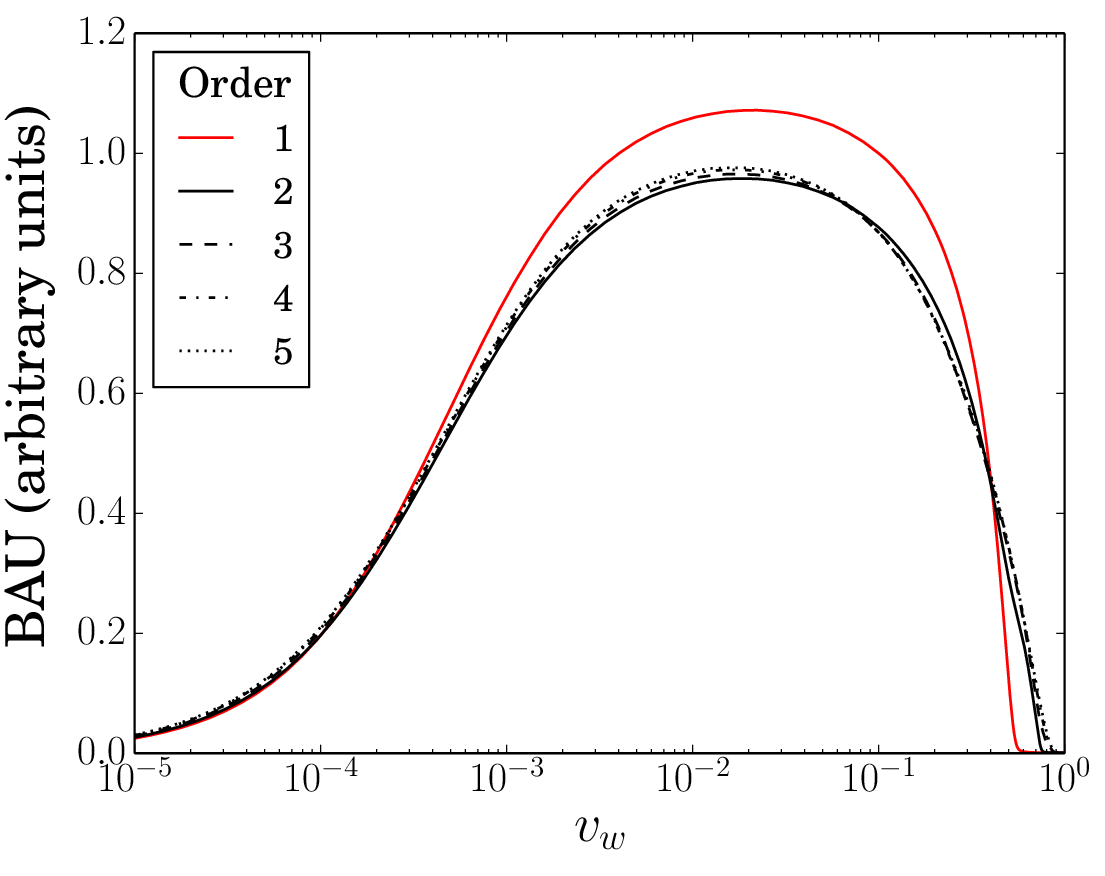}
	\caption{Same as figure~\ref{fig:results}, but now with a logarithmic scale in $v_w$ showing the behaviour at very small velocities.}
	\label{fig:smallvw}
\end{center}
\end{figure}

In figure~\ref{fig:smallvw} we extend the velocity dependence of the BAU down to very small values, $v_w\lesssim 0.001$. As expected, the result becomes highly suppressed and also converges rather quickly in this region. This is also in agreement with the findings in~\cite{Cline:2020jre}. We note also that the first-order approximation (with three perturbations, corresponding to the red curves) may be an over- or an underestimate of the full result, depending on the wall velocity and also on the value of the coupling in the collision terms.

\begin{figure}
\begin{center}
	\includegraphics[scale=.38]{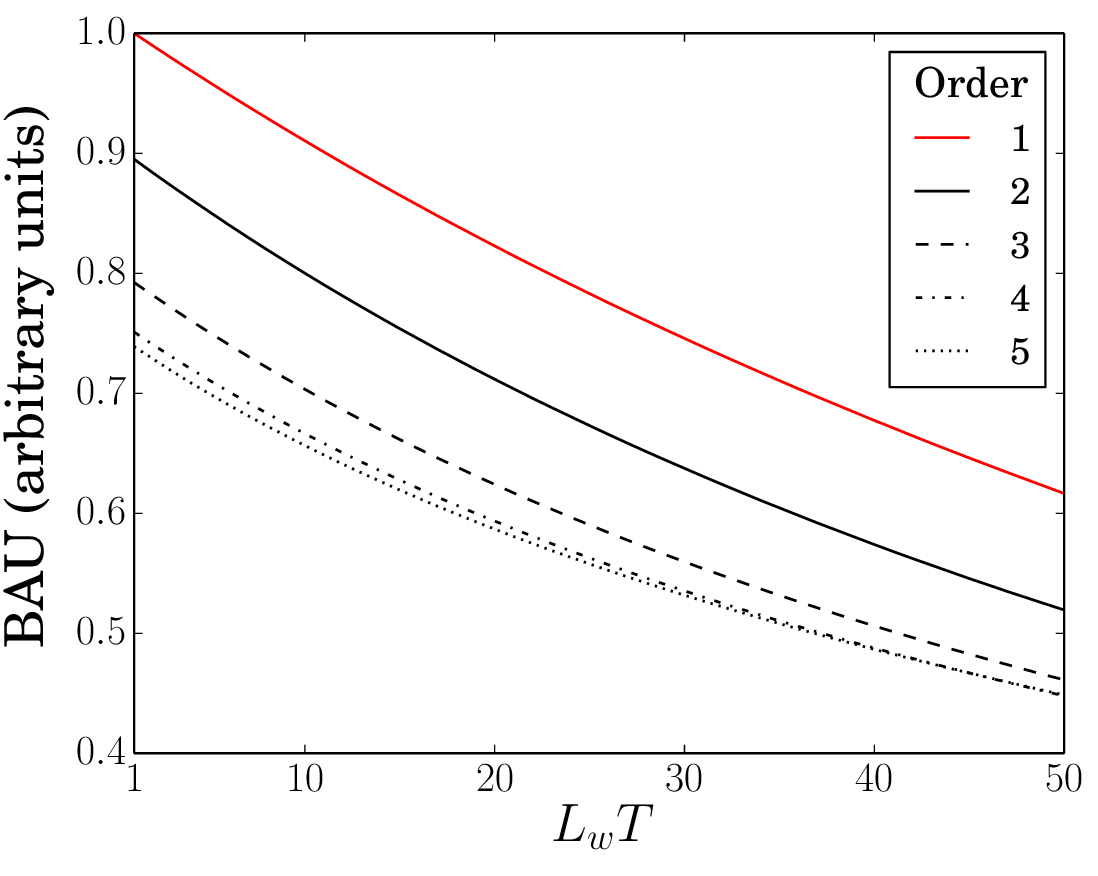}
	\caption{Dependence of the baryon asymmetry on the wall width for $\alpha_s=0.01$ and $v_w=0.4$.}
	\label{fig:Lw}
\end{center}
\end{figure}

Finally, the dependence with the wall thickness is displayed in figure~\ref{fig:Lw}. As one might expect, the shape of the curve does not change drastically by the addition of new fluctuations, but it is displaced as already seen in figures~\ref{fig:results} and \ref{fig:smallvw} above. The convergence of the series is also clearly highlighted in this plot.

\section{Conclusions}
\label{sec:conclusions}

In order to analyze electroweak baryogenesis or the bubble wall friction during a cosmological phase transition, 
a Boltzmann equation has to be solved for the particles in the plasma. Since a full solution of this 
system of partial differential equations is often (even numerically) not attainable, the main way to progress is to
take moments which leads to the Boltzmann hierarchy.

Some assumptions have to be made to decouple the Boltzmann hierarchy and the evaluation of the 
collision terms basically forces one to choose an Ansatz for the distribution functions. It turns out that 
if only a few moments and a few basis elements are chosen, the outcome critically depends on these choices.
For some choices, the speed of sound is an important quantity and baryogenesis for wall velocities 
beyond the speed of sound is insufficient, for others not. 

In this work, we use extensive basis sets (that generalize the fluid approximation) and many moments of the Boltzmann equation to study a toy model
that resembles the most important features of electroweak baryogenesis calculations. 
The main stumbling block for this kind of analysis is the evaluation of the collision term that we detailed in Appendix~\ref{sec:coll}. 
We find that for the fluid Ansatz and a low number of moments, supersonic baryogenesis is indeed suppressed.   
Then again, for a large number of moments, we find that the outcome behaves smoothly in wall velocity and supersonic 
baryogenesis becomes possible, in support of recent findings by Cline and Kainulainen~\cite{Cline:2020jre}.

The reason for the suppression in case of a few moments beyond the speed of sound is that the eigenvalues in the Liouville
operator change sign. For large wall velocities all eigenvalues share the same sign, fluctuations only exist behind the bubble
wall and baryogenesis becomes local. Qualitatively, it is the same for higher moments, but the velocity at which all eigenvalues 
share the same sign progressively moves to $v_w \to 1$. This is not too surprising since for a time-like wall velocity vector $v_\mu$
all fluctuations have to be damped in the positive time direction. So it is not too surprising that all eigenvalues have the 
same sign when $v_\mu$ approaches the light cone.

In order to leverage our results in a realistic baryogenesis calculation some ingredients are still missing. The main improvement would be to determine how the strong and weak sphalerons couple to the fluctuations in this generalized Ansatz. These are essential in the baryogenesis calculations since they break chirality and 
$B+L$ number. While we model these effect in our toy setup, representing these processes in a full analysis would require the evaluation of the corresponding collision/interaction rates 
for the extended fluid Ansatz.

Another important issue concerns the systematics governing the momentum expansion in the fluid Ansatz, and the criteria for deciding the appropriate order for truncation. {On general grounds one can expect the fluid to behave as nearly perfect if the mean free path is much smaller than the relevant macroscopic parameter, namely the wall width $L_w$. This indicates an expansion parameter of the order $(\Gamma L_w)^{-1}$, with $\Gamma$ denoting some combination of the relevant interaction ratios. This is in line with our findings that the convergence of the BAU gets better as the coupling constant (hence the interaction rates) become larger. As the mean free path increases, non-perfect behaviour such as viscosity and conductivity starts to become relevant, which are modelled by higher moments in the distribution function. All that said, it should be emphasized that a rigorous and a prioristic proof of the convergence of this moment expansion, or of the accuracy of any specific truncation, is still lacking~\cite{DeGroot:1980dk}. However, the convergent behaviour in numerous applications of this method, including the results of this work, corroborate the validity of this approach as long as non-linear phenomena (such as shock waves) are not present. A related issue is the accuracy of a fluid-like Ansatz in describing intermediate and low energy collisions. An expansion in powers of $p^\mu$ will naturally be dominated by the high energy regime, but may be less sensitive to complex angular behaviour of collisions at lower energies. This is not a problem in the leading-log approximation considered here, but may become an issue when attempting a leading order estimate. In this case a different Ansatz may become necessary, perhaps expanding in the energy $E$ and momentum direction $p^\mu/E$.}
Be it as it may, our numerical results show that, to leading-log accuracy, the conventional fluid approximation with three fluctuations seems to work fine at subsonic wall speeds when aiming at an $\mathcal{O}(10\%)$ accuracy. 

Finally, we comment how the present calculation relates to the evaluation of the bubble wall friction. Unlike the baryogenesis calculation, 
the friction calculation leads to a change in the collective temperature and velocity of the fluid through energy injection. This is captured by fluctuations 
in a background field that also become singular when the Liouville operator develops zero eigenvalues, {leading to a blow-up in friction as explicitly shown in reference~\cite{Konstandin:2014zta}}. In the baryogenesis calculation, this singularity in 
the source is counteracted by a divergent damping and no singularities occur in the final baryon asymmetry. {In the friction calculation, however , such divergences can be expected on the ground that there is a build up of particles in front of the wall precisely at the speed of sound, similar to a sonic boom effect, which drastically increases the friction in this case. In~\cite{Laurent:2020gpg} the behaviour of friction was analysed in the framework laid out in reference~\cite{Cline:2020jre}, and because the speed of sound plays no role in such formalism the findings point to a continuous non-singular behaviour for the friction at all values for the wall velocity. However, in our generalized fluid Ansatz with higher moments it is well 
possible that the singularities remain and yet others appear}, indicating that, in the limit of very high moments, the problem essentially becomes non-linear for wall velocities beyond the speed of sound.

\section*{Acknowledgements}
We thank B.~Laurent for helpful discussions.
GCD would like to acknowledge the support from Pr\'o-Reitoria de Pesquisa of Universidade Federal de Minas Gerais (UFMG) under grant number 28359*33.
 SJH is supported in part by the UK Science and Technology Facilities Council under Consolidated Grant ST/T00102X/1.
TK is supported by the Deutsche Forschungsgemeinschaft (DFG, German Research Foundation) under Germany`s Excellence Strategy -- EXC 2121 ``Quantum Universe'' -- 390833306.

\appendix

\section{An argument from hydrodynamics}
\label{sec:hydro}

Even though hydrodynamics does not describe the phenomenon of diffusion, the hydrodynamic analysis of bubble nucleation and expansion can shed some light on the relevance of the speed of sound. This is an old topic already presented in textbooks~\cite{Landau1987Fluid} and we will only summarize the main relevant points to the present discussion. 

While the wall is expanding, the interactions of the particles in the plasma with the Higgs field will drive the system out-of-equilibrium. This will lead to the fact that the equilibrium attained before and behind the wall are not the same. On length scales much larger than the bubble wall or diffusion length but smaller than the bubble size, the system is in local equilibrium. The energy-momentum tensor of the combined system (plasma+Higgs) is conserved and  the system can be described by hydrodynamics. Moreover, since there is no inherent length scale, the system behaves self-similarly and the hydrodynamic equations read~\cite{Espinosa:2010hh} 
\bea
\label{eq:hydro}
(\xi - v) \frac{\partial_\xi e}{w} &=& 2 \frac{v}{\xi} + \gamma^2 (1 - \xi v)\partial_\xi v \, ,\\ 
(1 - \xi v) \frac{\partial_\xi p}{w} &=& \gamma^2 (\xi - v) \partial_\xi v \, ,
\eea
where $v$ is the fluid velocity in the plasma frame, $\gamma$ is the corresponding Lorentz factor, $e$ and $p$ are the energy and pressure densities and $\xi$ is the self-similar coordinate $\xi=r/t$.

At the same time, in the vicinity of the wall, the Higgs field injects energy (and pressure) into the plasma which leads to the matching equations
\bea
\frac{v_+}{v_-}&=&\frac{e_b(T_-)+p_s (T_+)}{e_s (T_+)+p_b (T_-)} \, , \\
    v_+ v_-&=& \frac{p_s(T_+)-p_b(T_-)}{e_s(T_+)-e_b(T_-)} \, , 
\eea
where $u_\pm$ denote the plasma velocities (in the wall frame) and $T_\pm$ the temperature in front/behind the wall.

The interesting point about these equations is that only certain classes are valid globally. For example, 
if the wall velocity is supersonic (and the phase transition is not too strong), the fluid velocity in front of the wall has to vanish, since otherwise (\ref{eq:hydro}) would imply a singularity in the fluid velocities somewhere in front of the wall. So the two matching equations abide to $v_+=v_w$ and $T_+$ is the phase transition temperature. This leads to detonations ($v_-<v_+$) with a rarefaction wave developing behind the wall. 

On the other hand, if the wall velocity is subsonic, the fluid velocity behind the wall has to vanish since otherwise the solution will encounter a singularity according to (\ref{eq:hydro}). So the only valid solution in this regime is that of a shock before of the wall which are called deflagrations ($v_->v_+$). 
The fluid will then drop to the equilibrium configuration in the shock front (see ref.~\cite{Espinosa:2010hh} for details).

So, these solutions fulfill all the naive expectations. For supersonic wall velocities, the plasma changes only behind the wall while it is still in equilibrium in front of the wall. For subsonic wall velocities, particles have to be reflected which leads to a snow-plow effect that ultimately will build the shock in  front of the wall. Qualitatively, the solutions for subsonic and supersonic walls behave quite differently and one would expect that this also has to be reflected by the behavior close to the wall once one looks into the details of diffusion. 

Actually, there is also an intermediate regime where so-called hybrids develop with rarefaction waves and a shock. However, for very weak phase transition, this case becomes less and less relevant. Moreover, the strongest detonations (with the smallest wall velocity) are so-called Jouguet detonations. In this case, the fluid profile becomes very steep behind the wall ($dv/d\xi \to \infty$) due to the fact that the local fluid velocity (in the wall frame) is the speed of sound, so the Mach number is 1. 

The main takeaway from this section is that the macroscopic behavior of the fluid indeed changes from 
subsonic to supersonic wall velocities, which suggests that also microscopically the picture has to change qualitatively in this transition. Still, it is only very indirect evidence that baryogenesis is not possible for supersonic walls.

\section{Collision terms}
\label{sec:coll}

As explained in the last section, we will use the collision 
terms of the friction network to mimic the damping from the 
strong sphaleron in baryogenesis. We follow closely the analysis in~\cite{Moore:1995si} and also use for most parts their notation and conventions. The collision integrals are 
of the form\footnote{Compared to the notation in equation~(\ref{eq:Boltzmann}) one has $E_p C[f] = \mathcal{C}[\delta f]$.}
\be
C[f] = \sum \frac{1}{2E_p} \int_k \int_{p'} \int_{k'} 
|{\cal M}|^2 (2\pi)^4 \delta^4(p+k-p'-k') {\cal P}[f_i] \, ,
\ee
with
\be
{\cal P}[f_i] = f_p f_k (1 \pm f_{p'}) (1 \pm f_{k'}) -f_{p'} f_{k'} (1 \pm f_p) (1 \pm f_k) \, ,
\ee
and the shorthand $\int_k = \int d^3k/2E_k$.
The incoming momenta are denoted $k$ and $p$ and the outgoing $k'$ and $p'$. ${\cal M}$ is the matrix element of the process and $f_i$ are the particle distribution functions of 
the particles (that are fermionic/bosonic depending on process). 

We will evaluate the collision terms to linear order in the fluctuations (see~\cite{Moore:1995si}), i.e.
\be
{\cal P}[f_i] \simeq \left( \sum \delta_i \right) f^{eq}_p f^{eq}_k (1 \pm f^{eq}_{p'}) (1 \pm f^{eq}_{k'}) \, ,
\ee
where 
\be
f_p  = \frac{1}{\exp(p^\mu u_\mu/T + \delta) \pm 1} 
\simeq \delta \times (f_p^{eq})^\prime
\ee
and
\be
1 \pm f_p^{eq}  = \exp(p^\mu u_\mu/T) f_p^{eq}  \, .
\ee
Notice that this relation together with energy-momentum conservation in the process
implies that the last factor in ${\cal P}[f_i]$ is actually symmetric under exchange of 
$k$ with $p$, exchange of $k'$ with $p'$ and exchange of $k$,$p$ with $k'$,$p'$ as long as the statistics 
of these particles are the same. In effect the symmetries of ${\cal P}[f_i]$ depend then on the 
first factor $\sum \delta_i$.

Following~\cite{Moore:1995si} we are only interested in contributions that are logarithmically enhanced due to IR sensitivity, which only can arise from the 
$t$ and $u$ channels. The mass dependence of the particles  in the matrix element regulate the IR sensitivity of these integrals but we assume the particles to be massless otherwise. There are two types of 
diagrams we need to evaluate: annihilation diagrams and scattering diagrams.  
For the incoming particle with momentum $p$ we consider only fermions (quarks) and the scattering can happen off gauge bosons
or other fermions. 

Our Ansatz for the fluctuations we call collectively $\chi$ and expand 
\be
\chi(p) = \sum_a \chi_a(p) = 
 w^{(0)} + p^\mu w^{(1)}_\mu + p^\mu p^\nu w^{(2)}_{\mu\nu} + \cdots 
\ee
with
\bea
w^{(1)}_\mu &=& w^{(1)}_0 u_\mu + w^{(1)}_1 \bar u_\mu  \, ,\nn \\
w^{(2)}_{\mu\nu} &=& w^{(2)}_0 u_\mu u_\nu +  w^{(2)}_1 \bar u_\mu u_\nu +  w^{(2)}_2 \bar u_\mu \bar u_\nu \, ,
\eea
and so on.

In order to obtain the various moments of the Boltzmann equation, we multiply the Boltzmann
equation with some factors $\chi_a(p)$ and integrate over $p$. The outcome is 
\bea
\int d^3p \, \chi_a(p) C[f] &=& 
\int_p \int_k \int_{p'} \int_{k'} 
|{\cal M}|^2 (2\pi)^4 \delta^4(p+k-p'-k') \nn \\
&& \times \, \chi_a(p) \left( \sum \delta_i \right) f^{eq}_p f^{eq}_k (1 \pm f^{eq}_{p'}) (1 \pm f^{eq}_{k'})\, .
\eea

\begin{figure}
	\centering
	\includegraphics[page=1, scale=1.2]{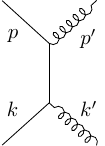}\qquad
	\includegraphics[page=2, scale=1.2]{plots/diagrams}\qquad
	\includegraphics[page=3, scale=1.2]{plots/diagrams}
	\caption{Diagrams for annihilation and scattering processes involving top quarks. Left: Annihilation into gluons in the $t$-channel. The $u$-channel is obtained by exchanging $p^\prime \leftrightarrow k^\prime$. Middle: Top quark scattering by other quark. Right: Scattering by gluons.}
	\label{fig:diagrams}
\end{figure}
\vskip 0.3cm
{\bf Annihilations:} The fluctuations are in the incoming particles while the produced particles are 
assumed to be in equilibrium. In case of the friction analysis the 
dominant contributions are CP and C conserving such that the incoming quarks and anti-quarks 
share the same fluctuations -- this is where a baryogenesis calculation would depart. The resulting 
integrals are then of the form
\bea
&& \int_p \int_k \int_{p'} \int_{k'} 
|{\cal M}|^2 (2\pi)^4 \delta^4(p+k-p'-k') \nn \\
&& \quad \times \, \chi_a(p) \left( \chi_b(p)  + \chi_b(k) \right) f^{eq}_p f^{eq}_k (1 \pm f^{eq}_{p'}) (1 \pm f^{eq}_{k'}) \, .
\label{eq:intAnn}
\eea
The matrix element for annihilations is  (we use the QCD rates of the Standard Model)
\be
|{\cal M}|^2 \simeq -\frac{64}{9} g_s^4 \frac{st}{(t - m^2_q)^2} \, .
\ee
We can symmetrize the expression by exchanging simultaneously $p$ with $k$ and $p'$ with $k'$
what essentially gives
\be
\chi_a(p) \left( \chi_b(p)  + \chi_b(k) \right) \to
\frac12 (\chi_a(p) + \chi_a(k)) \left( \chi_b(p)  + \chi_b(k) \right)
\ee
Remember that the factors like $\chi(p)$ will contain some products of the momentum with 
an tensor structure for the momenta. This will make the evaluation of these integrals 
somewhat cumbersome. Below we discuss three methods to deal with this Lorentz structure.

\vskip 0.3cm

{\bf Scattering processes:} We can follow the same steps and obtain expression like 
\bea
&& \int_p \int_k \int_{p'} \int_{k'} 
|{\cal M}|^2 (2\pi)^4 \delta^4(p+k-p'-k') \nn \\
&& \quad \times \, \chi_a(p) \left( \chi_b(p)  - \chi_b(p') \right) f^{eq}_p f^{eq}_k (1 \pm f^{eq}_{p'}) (1 \pm f^{eq}_{k'}) \, ,
\label{eq:intScatt}
\eea
where the matrix element is 
\be
|{\cal M}|^2 \simeq \frac{160}{3} g_s^4 \frac{s^2}{(t - m^2_q)^2}
\ee
for scattering off a quark and 
\be
|{\cal M}|^2 \simeq 40 g_s^4 \frac{s^2}{(t - m^2_q)^2}
\ee
for scattering off a gluon. Symmetrisation leads to the replacement
\be
\chi_a(p) \left( \chi_b(p)  - \chi_b(p') \right) \to
\frac12 (\chi_a(p) - \chi_a(p')) \left( \chi_b(p)  - \chi_b(p') \right) \, .
\label{eq:symmAnn}
\ee

\vskip 0.3cm

In the following we discuss three different ways to solve these integrals. The first one uses a 
symbolic solver to resolve and invert the Lorentz structure of the integrals. The second 
method is to evaluate the integrals numerically. The last one is to 
evaluate all integrals explicitly.

\subsection{Symbolic solver}

In this subsection we explain how to resolve the Lorentz structure arising from the basis
 functions $\chi_a(p)$ using a symbolic solver. We wrote a python code to automatically follow these steps. At some point the procedure becomes to 
expensive but we obtained results up to fourth order that we compared with ref.~\cite{Moore:1995si} as well as with
the explicit solution found below.

As noted in~\cite{Moore:1995si}, one can further simplify the integrals in (\ref{eq:intScatt}) and (\ref{eq:intAnn}) by only capturing the leading log. 
In particular, in the limit $t\to 0$ the difference ($p^\mu - p^{\prime\mu}$)
and ($k^\mu - k^{\prime\mu}$) are of order $\sqrt{t}$. This means that since the Matrix elements 
in combination with the basis functions $\chi$ behave as $1/t$, one can send 
$k'^{\mu} \to k$ and $p'^{\mu} \to p$ in the remainder of the integrals. Notice that 
this is for the scattering integrals only possible after the symmetrization (\ref{eq:symmAnn}).

\vskip 0.3cm
{\bf Annihilations:} 
At this stage, the only dependence on the momenta $p'$ and $k'$ appears 
in the matrix element and the Dirac delta function. Without loss of generality one
can boost into the frame where $\vec p + \vec k = 0$, and integrate over the spatial 
part of the delta function. This enforces $\vec p' + \vec k' = 0$. Since we assume the particles 
to be massless (which implies~$E_{k'} = E_{p'} = |\vec p'| = |\vec k'| = p'$ and $E_{k} = E_{p} = |\vec p| = |\vec k| = p$), the remaining integral is of the form (see~\cite{Moore:1995si})
\be
\int \frac{{p'}^2 dp' d\Omega_{p'}}{(2\pi)^3 2 E_{p'} 2E_{k'}} 
2\pi \delta(2E_p - 2E_{p'})
\frac{(2p_\mu k^\mu) 2pp'(1-\cos\theta')}{(2pp'(1-\cos\theta')+m_q^2)^2}
=\frac{1}{8\pi} \log \left(\frac{2 p^\mu k_\mu}{m_q^2}\right) \, .
\label{eq:ann_com}
\ee
Here $\cos\theta'$ denotes the angle between $\vec p$ and $\vec{p}^{\, \prime}$ and the final result 
was written in a Lorentz invariant way using $p_\mu k^\mu = 2pk$, which holds in the center-of-mass frame. 

The next step is to evaluate the Lorentz structure. Ultimately, the integrals only depend parametrically on the vector $u^\mu$, so the final result can only involve the vector $u^\mu$ and the metric $\eta_{\mu\nu}$. We construct the most general basis that has the correct symmetries and is build from these two ingredients. We then contract the Ansatz as well as the integral with the different basis elements and invert the system.

The resulting integrals can be evaluated in the plasma frame. The expression then involves the 
energies $E_p = p$ and $E_k = k$ as well as the Mandelstam variable $s = 2kp(1-\cos\theta)$, where $\cos\theta$
parametrizes the angel between $\vec k$ and $\vec p$. The integral over the angle reads 
\bea
&& \hskip -2cm
8\pi^2 \int d\cos\theta \, (1-\cos\theta)^n \,  
\log \left(\frac{2 kp (1-\cos\theta)}{m_q^2}\right) \nn \\
&=& \frac{2^{(n+1)}}{n+1} \left[\log \left(\frac{4 kp}{m_q^2}\right) - \frac{1}{n+1}\right] \nn \\
&\simeq& \frac{2^{(n+1)}}{n+1} \, \log \left(\frac{4 kp}{m_q^2}\right)
\eea
in leading log approximation. 

The remaining integrals factorize and using the approximation~\cite{Moore:1995si}
\be
\int dp \, p^n \log\frac{p}{T} f_p (1\pm f_p) \simeq 
\log(n+1/2)\int dp \, p^n  f_p (1\pm f_p) \, ,
\label{eq:MPapprox}
\ee
and the final integrals can be evaluated depending on the statistics of the particles, yielding the coefficients in equations~(\ref{eq:p_ints}) and (\ref{eq:ctilde}).
%
%

\vskip 0.3cm
{\bf Scattering:} The scattering diagrams are a bit harder to determine. The basis functions $\chi$ depend also on $p'$ such that the integrals over $p'$ and $k'$ are non-trivial. Still, also this problem can be solved by choosing an appropriate basis and inverting the problem by taking contractions of the integrals with this basis. 

Consider a term that contains the following Lorentz structure in the numerator,
\be
(p^\mu p^\nu \cdots - p'^\mu p'^\nu) (p^\alpha p^\beta \cdots - p'^\alpha p'^\beta) \,.
\label{eq:scattEx}
\ee
The most general outcome after integrating $p'$ and $k'$ can contain a tensor structure involving $k^\mu$, $p^\mu$ and the metric $\eta^{\mu\nu}$. Consider the contractions of this basis with terms of the form (\ref{eq:scattEx}). The contraction is at least of order $t$ such that the overall integrand scales as $t^{-1}$ just as the annihilation contributions. The Mandelstam variables are given by
\be
2 k^\mu p_\mu = s \, , \quad
2 k^\mu p'_\mu = -u = s + t \, , \quad
2 p^\mu p'_\mu = -t\, . \quad
\ee
First consider basis elements that contain factors $p$ and $k$ but not $\eta$. The corresponding contractions with (\ref{eq:scattEx}) are at least of order $t^2$ and can be discarded. Basis elements with more than one $\eta$ are also higher order in $t$. The only contributions linear in $t$ arise from basis elements with one factor $\eta$ and the remaining factors are all~$k$. We call the set of these basis elements $C$.

To invert the problem, one has to consider again the most general basis elements involving $k$, $p$ and $\eta$, and considers the class of basis elements that can give a non-vanishing contraction with elements of $C$ (meaning not suppressed by any factors $t$). One obvious class are elements with one factor $\eta$ and the remaining factors are $p$. But there is a second class: elements with one factor $k$ and the remaining factors are involve only $p$. 

The fundamental integrals over $p'$ and $k'$ that have to be solved are then of the same form as for annihilation processes and the resulting expressions have a Lorentz structure in $k$ and $p$ that can be resolved in the same way as for annihilations. One novelty is that the scattering on gauge bosons
involves integrals with two bosonic distribution functions as given in (\ref{eq:p_ints}).

\subsection{Numerical integration}

Another method to evaluate the collision integrals is to do it numerically, preferably with a Monte-Carlo integrator like CUBA~\cite{Hahn:2004fe}. One might think that this even allows to evaluate the integrals beyond the 
leading log approximation, but this is a mirage since the true result going beyond leading log requires 
to incorporate {\em hard thermal loop} correction in the scattering processes \cite{Arnold:2000dr,Arnold:2003zc,Wang:2020zlf}.

The only nontrivial problem in the numerical evaluation is how to represent the four-dimensional Dirac delta 
function. We choose to align the wall along the $z$-axis and then sample the six momenta $\vec k$ and $\vec p$. The vectors $\vec k'$ and $\vec p^{\,\prime}$ are then parameterized as 
\be
\vec k' = \vec P + \vec{Q}\, , \quad 
\vec p^{\,\prime} = \vec P - \vec{Q}\, , 
\ee
where $\vec P = (\vec k + \vec p)/2$ and we choose $\vec Q = q \, \hat q =  \,(0, q \sin\alpha,q \cos\alpha)$.
The remaining constraint on the energy then gives
\be
q^2 = \frac{\bar E^2  - \vec P^2}{\bar E^2 - (\hat q \cdot \vec P)^2} \bar E^2
\ee
with $2 \bar E = E_k + E_p$. Due to the symmetries of the integral and since we sample all signs of the six momenta $\vec k$ and $\vec p$, one can restrict oneself to the positive branch of $q$. Notice that 
this construction leads to an additional factor from the Jacobian determinant when the integral of the delta function is performed,
\be
\left| \frac{d(E_k + E_q + E_{k'} + E_{p'})}{dq} \right|^{-1}
= \left| \frac{q + (\hat q \cdot \vec P)}{E_{k'}}
+ \frac{q - (\hat q \cdot \vec P)}{E_{p'}} \right|^{-1}
\ee

We checked that the leading log result corresponds to the one obtained with the other two methods. 
The full numerical result can differ from the leading log up to a factor 2 in extreme cases (depending
on the involved masses that regulate the IR behavior). We checked that this 
has no effect on our main results.

\subsection{Explicit solutions}

It turns out that, to leading log order, the collision integrals for annihilation and scattering can be solved analytically to a simple closed form.

\vskip3mm

\textbf{Annihilations:} As discussed above, for annihilations one can always perform the $p^\prime$ and $k^\prime$ integrals in the center-of-mass frame, which from equation~(\ref{eq:ann_com}) results in

\be\begin{split}
	\int d^3p \, \chi_a(p) C[f] = \frac{16 \alpha_s^2}{9\pi^2} \sum_b\int_p \int_k \, &\chi_a(p) \,\big[ \chi_b(p) + \chi_b(k) \big]\, \times\\
		& \quad\times f^{eq}_p f^{eq}_k (1 + f^{eq}_{p}) (1 + f^{eq}_{k}) \log\left(\frac{2p^\mu k_\mu}{m_q^2}\right).
\end{split}\ee

The terms $\chi_{a,b}(p)$ contain powers of $E_p$ and $p_z$, so that the problem actually reduces to the solution of integrals of the form
\be
	4\pi^2T^{m+n+r+s+4}\gamma^{mnrs}_\text{ann} = \int \frac{d^3 p\, d^3k}{2E_p\, 2E_k}
			 E_p^m\, p_z^{r}\, E_k^{n}\, k_z^{s}\, f^{eq}_{p} f^{eq}_{k}(1+ f^{eq}_{p}) (1+ f^{eq}_{k}) \log\left(\frac{2p\cdot k}{m_q^2}\right)
\ee
(the pre-factor was chosen for later convenience as well as to make the quantity $\gamma^{mnrs}_\text{ann}$ dimensionless). 

By symmetry the result vanishes unless $r$ and $s$ are even, and one finds
\be
	\gamma^{mnrs}_\text{ann}= 
		\left\{\begin{array}{cl}
			\dfrac{\widetilde{c}_{m+r+2}\,\widetilde{c}_{n+s+2}}
				{(r+1)(s+1)} 
				\log\left(\dfrac{(2m+2r+3)(2n+2s+3)\,T^2}{m_q^2}\right), & r~\text{and}~s~\text{even}\\
				0, & \text{otherwise,}
		\end{array}\right.
	\label{eq:gann}
\ee
with the coefficients $\widetilde{c}_n$ given in equation~(\ref{eq:ctilde}).

\vskip3mm

\textbf{Scatterings:} The scattering integrals are significantly more complicated, but can be done with the assistance of a good deal of patience and perseverance. In this case equation~(\ref{eq:intScatt}), after antisymmetrizing according to (\ref{eq:symmAnn}), reduce to integrals of the form
\be\begin{split}
	\frac{\pi}{4} T^{m+n+r+s+4}\gamma^{mnrs}_\text{scatt} = 
		&\int  \frac{d^3 p\, d^3k}{2E_p\, 2E_k} f_{0p} f_{0k}(1- f_{0p}) (1\pm f_{0k})
		\int \frac{d^3p^\prime\, d^3k^\prime}{(2\pi)^6\,2E_{p^\prime}\,2E_{k^\prime}}\times\\
		&
		\qquad\times\frac{(2p\cdot k)^2}{(2p\cdot p^\prime + m_g^2)^2}
		(2\pi)^4 \delta^4(p+k-p^\prime-k^\prime)\times\\
		&\qquad\times\frac{(E_p^m p_z^r - E_{p^\prime}^m p^{\prime\ r}_z)
		(E_p^n p_z^s - E_{p^\prime}^n p^{\prime\ s}_z)}{2}.
	\label{gamma_scat}
\end{split}\ee
Again it is convenient to perform the primed integrals in the c.o.m. frame, but now this operation is complicated by the presence of primed factors of $E_{p^\prime}$ and $p^\prime_z$ in the integrand. To change the reference frame, let $u^\mu$ be the plasma four-velocity, which in the plasma frame is $u^\mu=(1,0,0,0)$, and let $v^\mu=(0,0,0,1)$ be a unit four-vector in the $z$ direction in the plasma frame. Then we can replace
\be
	E^m_{p^\prime} p^{\prime\,r}_z \to (u\cdot p^\prime)^m (v\cdot p^\prime)^r
\ee
in the integrand above, and since these are now Lorentz invariant quantities the $k^\prime$ integral can be easily performed in the c.o.m. frame, yielding remaining integrals of the form
\be\begin{split}
	&\int \frac{d^3 p^\prime}{(2\pi)^3 4E^2_{p^\prime}}
				\frac{(2p\cdot k)^2}{(2p\cdot p^\prime + m_g^2)^2}
				(2\pi) \delta(2E_p-2E_{p^\prime})\, \frac{(u\cdot p^\prime)^n (v\cdot p^{\prime})^r}{2}.
	\label{int_Empzr}
\end{split}\ee	
The task is now to write and solve this integral in the c.o.m. frame. For this purpose, let $\theta$ be the angle between $\vec{p}$ and $\vec{p}^{\,\prime}$, and $\beta$ the angle between $\vec{u}$ and $\vec{p}$, as seen in the c.o.m. frame. If we setup a coordinate system where $z$ is along $\vec{p}$, and with an appropriate choice of remaining axes, we can write
\be\begin{split}
	\mathbf{u} &= |\mathbf{u}|\, (\cos\beta\, \mathbf{\hat{p}} + \sin\beta\,\mathbf{\hat{y}} ),\\
	\mathbf{v} &= |\mathbf{v}|\, (\cos\alpha\,\mathbf{\hat{p}} + \sin\alpha\,\sin\rho\,\mathbf{\hat{y}}
							+ \sin\alpha\,\cos\rho\,\mathbf{\hat{x}}),\\
	\mathbf{p}^{\,\prime} &= |\mathbf{p}^{\, \prime}|\, (\cos\theta\, \mathbf{\hat{p}} + \sin\theta\,\sin\varphi\, \mathbf{\hat{y}} + \sin\theta\cos\varphi \,\mathbf{\hat{x}}),
	\label{vectors}
\end{split}\ee
and
\be\begin{split}
	(u\cdot p^\prime)^n &= \big[ u^0 E_{p^\prime} - |\mathbf{u}||\mathbf{p}^\prime|(\cos\theta\cos\beta + \sin\theta\sin\beta \sin\varphi ) \big]^n,\\
	(v\cdot p^\prime)^r &= \big\{ v^0 E_{p^\prime} - |\mathbf{v}||\mathbf{p}^\prime| \big[\cos\theta\cos\alpha + \sin\theta\sin\alpha\, (\sin\rho \sin\varphi + \cos\rho\cos\varphi) \big ] \big\}^r.
\end{split}\ee
We can now perform a binomial expansion of these expressions and integrate over $\theta$ and $\varphi$, keeping only leading log terms via
\be
	\int_{0}^\pi d\theta\,\frac{\cos^n\theta\,\sin\theta}{\left[2|\mathbf{p}||\mathbf{p}^\prime|(1-\cos\theta)+m_g^2\right]^2} 
		\simeq -\frac{n}{4|\mathbf{p}||\mathbf{p}^\prime|}\,\log\left(\frac{4|\mathbf{p}||\mathbf{p}^\prime|}{m_g^2}+1\right)\left(1+\frac{m_g^2}{2|\mathbf{p}||\mathbf{p}^\prime|}\right)^{n-1}.
\ee
Many terms will drop out and some of the binomial expansions can be resummed, especially when we set $m_g\to 0$ in the second brackets above. After some combinatorics the integral in equation (\ref{int_Empzr}) can be put in the form
\be\begin{split}
	 &\,\dfrac{1}{8\pi}\frac{(2p\cdot k)^2}{16|\mathbf{p}|^4}
			\log\left(\frac{4|\mathbf{p}|^2}{m_g^2}\right)\,
			\times ( u\cdot p)^{n-2} (v\cdot p)^{r-2}\Biggl\{\\
	&\qquad ~~~
			n\,|\mathbf{u}|^2 |\mathbf{p}|^2
			(v\cdot p)^2
		\bigg[
			\dfrac{u^0}{|\mathbf{u}|}\cos\beta - \cos^2\beta
			+ \dfrac{(n-1)}{2} \sin^2\beta 
		\bigg]\\
	& \qquad 
		+ r |\mathbf{v}|^2 |\mathbf{p}|^2
		(u\cdot p)^2 
		\bigg[
			\dfrac{v^0}{|\mathbf{v}|}\cos\alpha - \cos^2\alpha
			+ \dfrac{(r-1)}{2} \sin^2\alpha 
		\bigg]\\
	& \qquad+ (n\,r) (u\cdot p ) (v\cdot p )
		\bigg[ (u^0 |\mathbf{p}|)\,(v^0 |\mathbf{p}|) - (|\mathbf{u}||\mathbf{p}|\cos\beta)\, (|\mathbf{v}||\mathbf{p}|\cos\alpha)\bigg] 
	\Biggr\}.
\end{split}\ee

But this is only one of the four terms contributing to $\gamma^{mnrs}_\text{scatt}$ in equation~(\ref{gamma_scat}). When we add the other terms, other cancellations will take place, and the remainder can be written in a simple form in the plasma frame, involving only the terms
\begin{eqnarray*}
	2|\mathbf{p}|^2|\mathbf{u}|^2 \sin^2\beta &=& (E_p E_k + \mathbf{p}\cdot \mathbf{k})_\text{pl}\, , \\
	2|\mathbf{p}|^2 |\mathbf{v}|^2 \sin^2\alpha &=& \frac{1}{3}(3E_p E_k - \mathbf{p}\cdot \mathbf{k})_\text{pl}\, ,\\
	2 |\mathbf{u}||\mathbf{p}|\cos\beta &=& -(E_p-E_k)_\text{pl}\, ,\\
	2|\mathbf{v}||\mathbf{p}|\cos\alpha &=& (p_z - k_z)_\text{pl}\,,\\
	2 u^0 |\mathbf{p}| &=& (E_p + E_k)_\text{pl}\, ,\\
	2 v^0 |\mathbf{p}| &=& -(p_z + k_z)_\text{pl}.
\end{eqnarray*}
So, going back to the plasma frame,
\be\begin{split}
	\int_{p^\prime}\int_{k^\prime}
		&\frac{(2p\cdot k)^2}{(2p\cdot p^\prime + m_g^2)^2}
		(2\pi)^4 \delta^4(p+k-p^\prime-k^\prime)\,
		\frac{(E_p^m p_z^r - E_{p^\prime}^m p^{\prime\, r}_z)
		(E_p^n p_z^s - E_{p^\prime}^n p^{\prime\,s}_z)}{2} \simeq\\
	& \simeq \dfrac{1}{16\pi}
			\log\left(\frac{2p\cdot k}{m_g^2}\right)
			\times \Biggl\{
				(m\cdot n)\,(-p_z)^{r+s} E_p^{m+n-2} (E_p E_k + \mathbf{p}\cdot \mathbf{k})\\
		&\qquad\qquad\qquad 
				+ (r\cdot s)\,(-p_z)^{r+s-2} E_p^{m+n} \frac{1}{3}(3E_p E_k - \mathbf{p}\cdot \mathbf{k})\\
		&\qquad\qquad\qquad
			-(ms+nr)( E_k p_z + E_p k_z) E_p^{m+n-1}  (-p_z)^{r+s-1}
	\Biggr\}_\text{pl}.
\end{split}\ee

Finally, this can be integrated over $p$ and $k$ to yield a vanishing result for $r+s$ odd, while for $r+s$ even one has
\be\begin{split}
	\gamma^{mnrs}_\text{scatt} =~&
		\log\left(\frac{5(2m+2n+2r+2s+1)T^2}{m_g^2}\right)\times\\
			&\quad\times\left( \frac{ms + m n + nr}{r+s+1} + \frac{r\,s}{r+s-1} \right) \, c^f_{m+n+r+s} 
			\left\{ \begin{array}{c} 
				c_2^f\\ 
				c_2^b
			\end{array}\right\}.
	\label{eq:gscat}
\end{split}\ee
The term $c_2^f$ enters in scatterings by fermions and $c_2^b$ appears in scatterings by gluons.


\end{document}